\documentclass[aps,pre,twocolumn,superscriptaddress,amsmath,amssymb]{revtex4}
\usepackage{amsmath}
\usepackage{epsfig}

\newcommand{\bra}[1]{\left\langle #1\right|}
\newcommand{\ket}[1]{\left|#1\right\rangle}
\newcommand{\braket}[2]{\left\langle #1 \left|#2\right.\right\rangle}

\begin{document}

\title{Thermodynamics of Coarse Grained Models of Super-Cooled
Liquids}

\author{David Chandler}

\affiliation{Department of Chemistry, University of California,
Berkeley, CA 94720-1460}

\author{Juan P.  Garrahan}

\affiliation{School of Physics and Astronomy, University of
Nottingham, Nottingham, NG7 2RD, UK}

\date{\today}

\begin{abstract}
In recent papers, we have argued that kinetically constrained coarse
grained models can be applied to understand dynamic properties of
glass forming materials, and we have used this approach in various
applications that appear to validate this view. In one such paper
[J.P. Garrahan and D.  Chandler, Proc. Nat. Acad. Sci. USA
\textbf{100}, 9710 (2003)], among other things we argued that this
approach also explains why the heat capacity discontinuity at the
glass transition is generally larger for fragile materials than for
strong materials. In the preceding article, Biroli, Bouchaud and
Tarjus (BB\&T) have objected to our explanation on this point, arguing
that the class of models we apply is inconsistent with both the
absolute size and temperature dependence of the experimental specific
heat. Their argument, however, neglects parameters associated with the
coarse graining.  Accounting for these parameters, we show here that
our treatment of dynamics is not inconsistent with heat capacity
discontinuities.
\end{abstract}

\maketitle

\section{Introduction}

Many workers argue that a thermodynamic anomaly underlies the onset of
glassy dynamics~\cite{Thermo}. Evidence for this view is the rough,
though not quantitative~\cite{NotQuant}, correlation between dynamic
fragility and excess heat capacity
discontinuity~\cite{Coincidence}. This thermodynamic view contrasts
with the picture we have advocated~\cite{JPGDC02,otherpapers},
attributing glassy behavior to dynamic heterogeneity~\cite{DH}, with
growing length scales appearing in space-time, but not space
alone~\cite{JPGDC02}.  Indeed, direct observations of diffusive motion
in colloidal glasses reveal excitations that are local and
sparse~\cite{Colloid}.  These findings seem consistent with
excitations being local and uncorrelated, as assumed in the two-state
model for the low temperature behavior of structural glass heat
capacities~\cite{Twostate}. Spatial correlations under such conditions
can arise through constraints on particle motions that are relieved
only when adjacent regions have exhibited some degree of
mobility~\cite{Glarum, FA}. Trajectories are then correlated
throughout space and time, with varying degrees of hierarchical
structure determining the extent to which the system is fragile or
strong~\cite{Palmer}. From this perspective, a non-thermodynamic
explanation emerges for correlation between heat capacity and
fragility~\cite{JPGDC03}: The concentration of excitations required to
support fragile hierarchical dynamics is higher than that required for
strong non-hierarchical dynamics. The juxtaposition of heat capacity
discontinuities is therefore understood as a consequence of different
excitation concentrations.

Biroli, Bouchaud and Tarjus (BB\&T)~\cite{BBT} have taken issue with
this explanation.  They show that the simplest treatment of defect
models cannot simultaneously fit the size and temperature dependence
of the experimental heat capacity while simultaneously fitting dynamic
properties. In general and in more current context, however, the
terminology ``defect models'' refers to a broad class of kinetically
constrained models~\cite{Ritort-Sollich}.  As a result of this
generality, their question ``Are defect models consistent with the
entropy and specific heat of glass-formers?'' has a non-trivial answer
(see also the analysis in
\cite{Angell-Rao,Moynihan-Angell,Matyushov-Angell}).  In particular,
this class of models has an assortment of possible dynamical
behaviors, so that relaxation data can be fit with many different
functions of excitation concentration. In addition, kinetically
constrained models are coarse grained caricatures of fluids, so that
many degrees of freedom remain unspecified. Here, we focus on this
latter feature, the possible consequences of unspecified degrees of
freedom.

The starting point is the assumption that small length-scale and small
time-scale features of a fluid can be integrated out leaving only
simple stochastic rules for the dynamics of discrete variables on a
lattice. These rules contain constraints imagined to be the
consequence of the actual intermolecular interactions, interactions
that limit the space or metric for molecular
motions~\cite{Metric}. The excitations or defects that survive coarse
graining distinguish microscopic regions of space-time that exhibit
molecular mobility from those where molecules are jammed or
immobile. In particular, an excitation or defect is a microscopic
region of space for which particle mobility emerges. This
characterization is related to coarse graining in time, because it
takes time to discern whether or not mobility is exhibited.  Bear in
mind, ``defect'' does not distinguish disordered arrangements of
atoms from those that are ordered because there are generally many
disordered yet jammed configurations. Similarly, a lack of
``excitation'' does not necessarily imply low energy because some
immobile regions may have the same energy as mobile
regions. Uncertainty concerning this terminology may be the origin of
criticisms of facilitated models leveled by BB\&T~\cite{BBT,Bouchaud}
and also by Lubchenko and Wolynes~\cite{Wolynes}.

As a simple illustration, consider dividing space into cells, with
grid spacing larger than the equilibrium correlation length of the
material. At a given time frame, the micro states of different cells
are then uncorrelated.  Further, suppose there are two energy levels,
$0$ and $J$, for the states of a given cell, where the states with
energy $J$ include those that will exhibit particle mobility after a
coarse graining time $\delta t$. There is a fraction, $\phi$, of
states with energy $J$ that exhibit mobility in this way; the others
remain jammed. In this case the average excitation concentration is
$\phi$ times the concentration of cells with energy $J$, and thus the
number of molecules contributing to energy or enthalpy fluctuations is
$\phi^{-1}$ times larger than that contributing to mobility
fluctuations. In other words, when considering the thermodynamic
implications of a kinetically constrained defect model, a factor
$\phi^{-1}$ is required to account for energetic states or degrees of
freedom absent from explicit consideration. BB\&T ignore this factor,
which leads them to the difficulties they describe.

In the next section we discuss this point in greater detail. We show
that facilitated or kinetically constrained models such as the one we
presented in Ref.~\cite{JPGDC03} admit a variety of possibilities for
partitioning micro states. This demonstration justifies the heat
capacity formulas we used in Ref.~\cite{JPGDC03} and therefore
supports our argument about the juxtaposition of heat capacities for
strong and fragile glass formers.

\section{Models}

The dynamics of facilitated models we consider are governed by
master equations for distribution functions of the mobility field on a
lattice. Consistent with our earlier papers, we use the symbol
$n_{i}(t)$ to denote the value of the mobility field at lattice site
$i$ during time period $t$. It is a binary field in that $n_i$ takes
on one of two values, $0$ (corresponding to a cell that is unexcited,
jammed) or $1$ (corresponding to excited, mobile). Each cell is
imagined to contain several molecules, so we expect that even after
some coarse graining, there are still many (not just two) microstates
possible for each cell at a given time $t$. We further imagine these
many microstates, specified with $\mu _{i}$, evolve according to their
own master equation. The master equation for the distribution of the
mobility field $n_{i}$ is a contraction of that for the micro-state
field $\mu _{i}$. In this section, we show how this contraction
works. We begin by considering the thermodynamics and statistics of
the fields $n_{i}$ and $\mu _{i}$, and then consider the master
equations for the distributions of these fields.

\subsection{Excitations and microstates}
\label{thermo}

Imagine we partition the liquid into a lattice with grid spacing
$\delta x$, and denote the state in cell $i$ by $\mu _{i}$.  Imagine
further that the set of available cell levels $\{ \mu_i \}$ can be
split into two groups: one subset of $\{ \mu_i \}$ is ``unexcited''
with respect to mobility, $n_i=0$, and the complementary subset is
``excited'', $n_i=1$.  In practice, $n_{i}$ is most conveniently
determined by observing system for a coarse graining time $\delta t$.
This splitting, or projection, of many micro states into two
different mobility states (or several, as in the generalization of
\cite{JPGDC03}) is the the first central assumption of facilitated
models \cite{Ritort-Sollich}.

While the state $\mu$ of a cell will correspond to one of the two
mobility states, $n=0,1$, in general its energy, $\epsilon_\mu$, will
be distributed.  (For notational simplicity, when clarity is not
diminished, we drop the subscripts on $n_i$ and $\mu_i$.) Let
$G_0(\epsilon)$ and $G_1(\epsilon)$ represent the normalized
probability densities for the energy levels in states $n=0$ and $n=1$,
respectively.  The cell partition function then reads (we set
$k_B=1$):
\begin{equation}
Z = \sum_\mu e^{-\epsilon_\mu / T} = z_0 + z_1 ,
\end{equation}
where
\begin{equation}
z_n \equiv \omega_n \int d\epsilon ~ G_n(\epsilon) e^{-\epsilon / T}
\;\;\; (n=0,1) ,
\end{equation}
and $\omega_n$ gives the total number of levels in each of the two
mobility states.  The excitation concentration is given by:
\begin{equation}
c \equiv \langle n \rangle = \frac{z_1/z_0}{1 + z_1/z_0} .
\end{equation}
The average energy per cell is:
\begin{equation}
\langle \epsilon \rangle = (1-c) \langle \epsilon_0 \rangle + c
\langle \epsilon_1 \rangle ,
\end{equation}
where
\begin{equation}
\langle \epsilon_n^k \rangle \equiv \frac{\int d\epsilon ~ \epsilon^k
G_n(\epsilon) e^{-\epsilon / T}}{\int d\epsilon ~ G_n(\epsilon)
e^{-\epsilon / T}} \;\;\; (n=0,1) .
\end{equation}
The specific heat per cell is:
\begin{equation}
C_v = T^{-2} \left( \langle \epsilon_1 \rangle - \langle \epsilon_0
\rangle \right)^2 c (1-c) + C_v^{(0)} (1-c) + C_v^{(1)} c ,
\end{equation}
where
\begin{equation}
T^2 C_v^{(n)} \equiv \langle \epsilon_n^2 \rangle - \langle \epsilon_n
\rangle^2 \;\;\; (n=0,1) .
\end{equation}

The quantity $\ln{(z_1/z_0)}$ determines the
concentration $c$ of excitations; $\langle \epsilon_1 \rangle -
\langle \epsilon_0 \rangle$ gives the average cell energy; and
$T^{-2} \left( \langle \epsilon_1 \rangle - \langle \epsilon_0
\rangle \right)^2 + C_v^{(1)} - C_v^{(0)}$ sets the specific heat at
small $c$.  Except in the case where $G_n(\epsilon)$ are delta
functions, these scales are in principle all different.

As a simple example, which will be useful below, consider the
following model with just two energy levels.  The unexcited states 
$n=0$ occupy both energy levels, which we
denote $g$ and $h$, with energies $\epsilon_g=0$ and
$\epsilon_h=J$, respectively:
\begin{equation}
G_0(\epsilon) = (1-\alpha) \delta(\epsilon) + \alpha
\delta(\epsilon-J) ,
\end{equation}
where $\alpha$ sets the relative degeneracy between the two levels.
The excited states $n=1$ have a single energy level, which we denote
$x$, with energy $\epsilon_x=J$:
\begin{equation}
G_1(\epsilon) = \delta(\epsilon-J) .
\end{equation}
In this case, the cell partition function, average energy and specific
heat become:
\begin{eqnarray}
Z &=& (1-\alpha) \omega_0 + e^{-J / T} \left( \alpha \omega_0 +
\omega_1 \right) \label{Z3state} \\ \langle \epsilon \rangle &=& J
(1-c) c_h + J c \\ C_v &=& (J/T)^2 \left( 1-c_h \right)^2 c (1-c)
\nonumber \\ && + (J/T)^2 c_h \left( 1-c_h \right) (1-c) ,
\end{eqnarray}
where $c_h$ denotes the relative occupation of the states $h$ with
respect to states $g$, $c_h \equiv \alpha e^{-J / T} /
[(1-\alpha)+\alpha e^{-J / T}]$.  The total concentration of levels
with energy $J$ is
\begin{equation}
c + (1-c) c_h \equiv \phi^{-1} ~ c ,
\label{defphi}
\end{equation}
where we have defined $\phi$ as the ratio of mobile cells to the total
number of cells with energy $J$.  The specific heat then reads
\begin{equation}
\label{cv3state}
C_v = (J/T)^2 ~ \phi^{-1} ~ c ~ (1 - \phi^{-1} ~ c) .
\end{equation}

\subsection{Kinetics}
\label{dyna}

We now show how an appropriate projection of the dynamics of the full
states $\mu_i$ of individual cells into their mobility states
$n_i=0,1$ gives a two-state facilitated model such as the FA or East
models, and the slightly more general class of models introduced in 
Ref.\cite{JPGDC03} .  The approach we take has been used to derive 
approximate real-space renormalization group 
equations~\cite{Whitelam-Garrahan-PRE}.  Here, however, the approach 
requires no approximation.

The master equation for the dynamics can be written \cite{Siggia}:
\begin{equation}
\frac{\partial}{\partial t} \ket{P(t)}= -\mathcal{L} \ket{P(t)},
\end{equation}
where $\ket{P(t)}$ is the state vector for the probability density of
the system at time $t$, and $\mathcal{L}$ is the Liouvillian operator
for the dynamics.  In the direct product representation it reads
\cite{Whitelam-Garrahan-PRE}:
\begin{equation}
\ket{P(t)} = \sum_{\{\mu\}} P(\mu_1 \cdots \mu_N,t) \, \ket{\mu_1}
\otimes \ket{\mu_2} \otimes \cdots \otimes \ket{\mu_N},
\end{equation}
where $\ket{\mu_i}$ is the state vector for level $\mu_i$ in site $i$.

The second assumption in facilitated models is that transitions in a
cell $i$ are only possible if neighbouring cells $j$ are mobile,
$n_j=1$ \cite{Ritort-Sollich}.  In a system with facilitated dynamics
$\mathcal{L}$ reads \cite{Whitelam-Garrahan-PRE}:
\begin{equation}
\mathcal{L} = \sum_i \mathcal{C}_i \ell_i
\end{equation}
where $\mathcal{C}_i$ is the kinetic constraint operator imposing
facilitation, and $\ell_i$ is the unconstrained dynamic operator at
site $i$.  For example, in the East facilitated model $\mathcal{C}_i =
n_{i-1}$, and in the direct product representation:
\begin{equation}
\mathcal{L} = \sum_{i} \mathbf{1} \otimes \cdots \otimes n_{i-1}
\otimes \ell_i \otimes \mathbf{1} \otimes \cdots \otimes \mathbf{1} .
\end{equation}
The off diagonal elements of $\ell_i$ are given by the transition
rates between levels: $(\ell_i)_{\mu' \mu} = - \gamma_{\mu \to \mu'}$,
while the diagonal elements are such that the sum of columns in
$\ell_i$ vanish.  A choice satisfying detailed balance is
$(\ell_i)_{\mu' \mu} = - \braket{\mu'}{\rm{eq}}$, where $\ket{\rm{eq}}$
is the equilibrium distribution.  Notice that we have assumed that
transitions between any of the levels in the cell are only possible if
facilitated by a neighbour.

In order to project the dynamics of the full set of levels $\ket{\mu}$
into that of only the excitation occupancies $\ket{n=0,1}$ we need
projection, $T_1$, and embedding, $T_2$, operators
\cite{Whitelam-Garrahan-PRE}:
\begin{equation}
\label{t1t2}
T_1 = \sum_{n,\mu} a_{n,\mu} \ket{n} \bra{\mu}, \;\;\; T_2 =
\sum_{\mu,n} b_{\mu,n} \ket{\mu} \bra{n} ,
\end{equation}
where the $a$ and $b$'s are constants.  The projection is performed
independently in each cell, so these are site diagonal operators.
They must also satisfy $T_1 T_2 = \mathbf{1}$.  The dynamics of the
projected state $\ket{P(t)}' \equiv T_1 \ket{P(t)}$ is then given by
the projected Liouvillian
\begin{equation}
\mathcal{L}' = T_1 ~ \mathcal{L} ~ T_2 ,
\end{equation}
where the elements of the projection and embedding matrices
(\ref{t1t2}) are given by:
\begin{equation} 
a_{n \mu} = \delta_{n, n_\mu} , \;\;\;
b_{\mu n} = \braket{\mu}{\rm{eq}} \left( \frac{1-n}{1-c}+\frac{n}{c}
\right)  \delta_{n_\mu, n} .
\end{equation} 
The choice of rates in $\mathcal{L}$, which depend only on the final
level, $\gamma_{\mu \to \mu'}=\braket{\mu'}{\rm{eq}}$, ensures that
the projected dynamics is Markovian \cite{Sharpe}.

We now show explicitly how the projection procedure works in the
simple the three-level example of the previous subsection.  In this
case $\ket{\mu} = \ket{g},\ket{h},\ket{x}$, and the onsite operators
can be represented by $3 \times 3$ matrices:
\begin{eqnarray}
n &=& \left(
\begin{array}{ccc}
0 & 0 & 0 \\ 0 & 0 & 0 \\ 0 & 0 & 1
\end{array} \right) , \\
\ell &=& \left(
\begin{array}{cc}
c+(1-c) c_h & -(1-c)(1-c_h) \\ -(1-c) c_h & c+(1-c) (1-c_h) \\ -c & -c
\end{array} \right.   \nonumber \\
&& \ \ \ \ \ \ \ \ \ \ \ \ \ \ \ \ \ \ \ \ \ \ \ \ \ \ \ \ \left.
\begin{array}{c}
  -(1-c)(1-c_h) \\ -(1-c) c_h\\ 1-c
\end{array} \right)
\end{eqnarray}

Of the three levels $\ket{\mu}$, only $\ket{x}$ corresponds to the
excited state, and the projection matrix is:
\begin{equation}
T_1 = \left(
\begin{array}{ccc}
1 & 1 & 0 \\ 0 & 0 & 1
\end{array} \right) .
\end{equation}
The corresponding embedding matrix is:
\begin{equation}
T_2 = \left(
\begin{array}{cc}
1-c_h & 0 \\ c_h & 0 \\ 0 & 1 \\
\end{array} \right) .
\end{equation}
The projected operators then read:
\begin{eqnarray}
n' &=& T_1 n T_2 = \left(
\begin{array}{cc}
0 & 0 \\ 0 & 1
\end{array} \right) , \\
\ell' &=& T_1 \ell T_2 = \left(
\begin{array}{cc}
c & c-1 \\ -c & 1-c
\end{array} \right) .
\end{eqnarray}
These are precisely the operators for a two state facilitated model
\cite{Whitelam-Garrahan-PRE}.

\begin{figure}
\begin{centering}
\includegraphics[width=8.6cm]{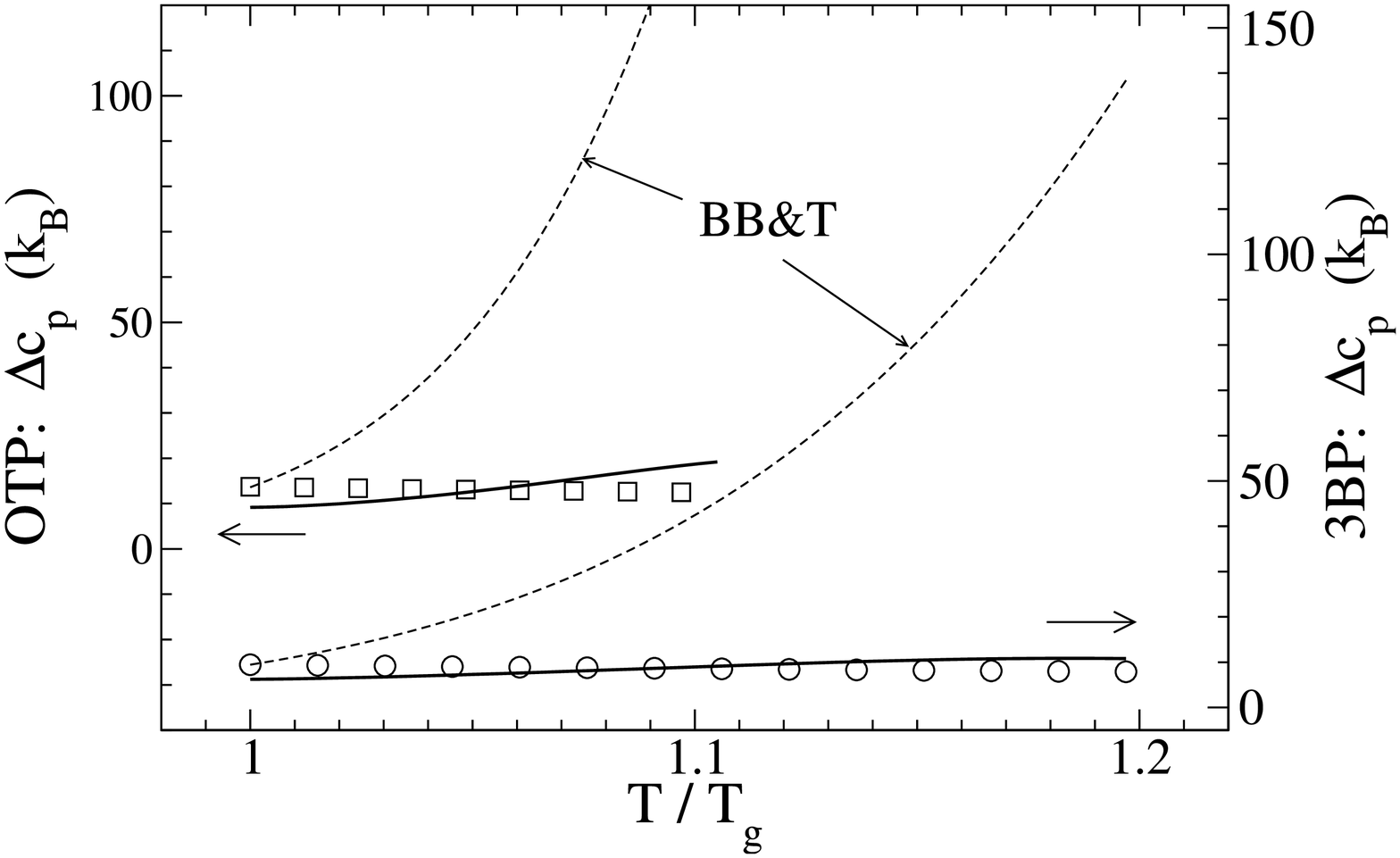}
\includegraphics[width=8cm]{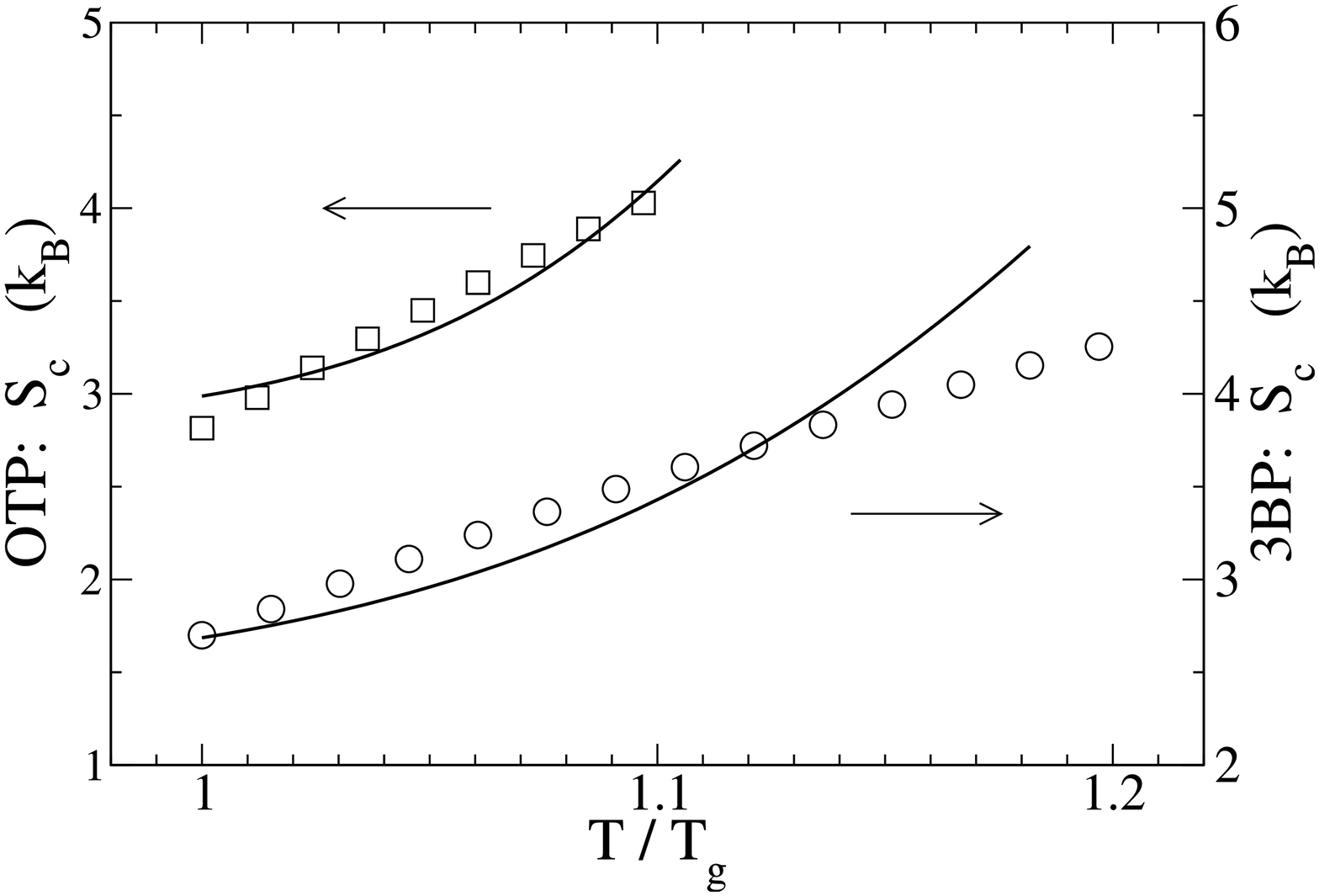}
\caption{Fitting of thermodynamic data near $T_g$ using the
facilitated models of Ref.\ \cite{JPGDC03} and compared to the formula
used by Biroli, Bouchaud and Tarjus (indicated as BB\&T) \cite{BBT}.
{\bf Top panel}: excess specific heat, $\Delta c_p$, of supercooled
OTP (left hand scale) and 3BP (right hand scale).  Symbols indicate
experimental data \cite{Moynihan-Angell}.  Full curves are fits using
the formulas for the heat capacity per molecule discussed in the text
(where $C_{\rm mol.} \equiv \Delta c_p$): $C_{\rm mol.} = \mathcal{N}
(J/T)^2 \phi^{-1} c (1 - \phi^{-1} c)$, where $\mathcal{N} = (\phi
s)^{-1}$ and $\phi = \phi_{\mathrm{g}}+\phi_{1}\left(
T-T_{\mathrm{g}}\right)/T_g$.  Here $c \approx \phi e^{-J/T} \omega_1
/ [(1-\alpha) \omega_0] \approx g \exp \left( -J/T +
J/T_{\mathrm{ref}} \right)$, where the first (approximate) equality
neglects terms of order $c^{2}$, and the second introduces the
notation used in our Ref.\ \cite{JPGDC03}: $g$ is a degeneracy that we
imagine measures the possible directions of mobility, and
$T_{\mathrm{ref}}$ bounds from above the temperatures we consider.
The values of $J \approx 26.7$ and $16.7$ for OTP and 3BP were
determined in \cite{JPGDC03} from kinetic data.  On physical grounds
we expect $s \approx (J/T_m)/\Delta S_{\rm fusion} \sim 1-10$, and
$\phi$ to increase with temperature.  The fitting of $\Delta c_p$
gives $s,\phi_g,\phi_1 = 5, .03, .7$ and $4.5, .04, .5$ for OTP and
3BP, respectively.  $\mathcal{N} \sim 6$ at $T_g$ as argued in
\cite{JPGDC03}.  Clearly, it is possible to describe the size and
modest change with $T$ of the specific heat in the vicinity of $T_g$,
in contrast to the observation of BB\&T.  {\bf Bottom panel}: excess
entropy, $S_{\rm c}$, of OTP (left hand scale) and 3BP (right hand
scale).  Symbols correspond to experimental data
\cite{Moynihan-Angell}.  Full curves are fits using the formula for
the entropy per molecule which follows from Eq.\ (\ref{C1}): $S_{\rm
mol.} = s^{-1} \{ \ln{[(1-\alpha) \omega_0]} - (\phi^{-1} c)
\ln{(\phi^{-1} c)} - (1- \phi^{-1} c) \ln{(1-\phi^{-1} c)} +
(\phi^{-1} c) \ln{[(\omega_0 \alpha + \omega_1)/(1-\alpha) \omega_0]}
\}$.  The shown fits correspond to the same parameters as before plus
$s^{-1} \ln{[(1-\alpha) \omega_0]}, s^{-1} \ln{[(\omega_0 \alpha +
\omega_1)/(1-\alpha) \omega_0]}=2.8, 1.2$ and $2.4, 1.8$ for OTP and
3BP, respectively.}
\end{centering}
\end{figure}

\section{Data fitting}

Subsection \ref{thermo} tells us that if the mobility field is the
result of coarse-graining and projecting the many mobile and immobile
micro states, then different energy (or enthalpy) scales enter into
the concentration of mobile cells $c$, the average energy (or
enthalpy), and the specific heat.  How these scales relate to each
other depends on the distribution of energy levels.  At equilibrium,
the average excitation concentration, $c$, is a function of
thermodynamic state.  Its value should be indicative of the
concentration of regions with excess energy or enthalpy, but to say
more requires more assumptions or detail than needed to predict
dynamical trends.

In order to illustrate how facilitated models can be used to fit
thermodynamic data the simple example given above of two immobile cell
levels $g$ and $h$ and a single mobile level $x$ will suffice.  From
Eqs.\ (\ref{Z3state})--(\ref{cv3state}) we see that while $c$ gives
the concentration of cells displaying mobility, both mobile and
immobile cells can contribute to energy (or enthalpy) fluctuations.
The relevant quantity to determine the specific heat is not $c$ but
$c+(1-c)c_h$.  From Eqs.\ (\ref{defphi}) and (\ref{cv3state}) we have
that the specific heat {\em per cell} is:
\begin{eqnarray}
C_{\rm cell} &\approx&\phi ^{-1}\left( J/T\right) ^{2}c\left( 1-\phi
^{-1}c\right) \, \nonumber \\ &\thicksim &\phi ^{-1}\left( J/T\right)
^{2}c, \label{C1}
\end{eqnarray}
The asymptotic equality of Eq.\ (\ref{C1}) is, in effect, the
expression we refer to in the penultimate paragraph of Ref.\
\cite{JPGDC03}.  The factor $\phi ^{-1}$ accounts for the number of
molecules contributing to energy or enthalpy fluctuations, this number
being larger than that for molecules contributing to mobility
fluctuations.  This fact is implicit in the coarse graining procedure,
as shown above.  In Ref.\ \cite{JPGDC03}, $\phi^{-1}$ is included in
the quantity we called $\mathcal{N}$.  In the few lines devoted to
this topic in Ref.~\cite{JPGDC03}, the meaning of $\mathcal{N}$ was
obscure and its description misleading.  Indeed, where we wrote
``$\mathcal{N}$ is the number of molecules that contribute to enthalpy
fluctuations per mobile cell'', we should have written ``$\mathcal{N}$
accounts for the number of molecules that contribute to enthalpy
fluctuations beyond those associated with mobile cells.''

To be perfectly explicit, if $s$ is the average number of molecules in
a cell, then $\mathcal{N}=1/\phi s$.  With this expression, the
relative values of $C$ for several super-cooled liquids at $T_g$ are
consistent with the relative values of their heat capacity jumps at
the glass transition.  This consistency is independent of the value of
$s$ and $\mathcal{N}$, but the value of $\mathcal{N}$ does play a role
in the actual value of the heat capacity discontinuity.  Specifically,
\begin{equation}
C_{\rm mol.} \approx s^{-1} C_{\rm cell} \approx \mathcal{N}\left(
J/T\right)^{2}c ,
\end{equation}
is consistent with the value of the discontinuities of the heat
capacity per molecule at $T_{\rm g}$ when $\mathcal{N}\approx 6$.
BB\&T \cite{BBT} identify $s$ with $\mathcal{N}$, and further neglect
the factor $1/\phi$, and in so doing disagree with this consistency.

BB\&T's second criticism is about the temperature variation of
$C$.  They consider the relationship
\begin{equation}
C_{\mathrm{BB\&T}} \propto \left( J/T\right)^{2} c .
\label{CbyBB&T}
\end{equation}
This formula is consistent with the asymptotic equality of Eq.\
(\ref{C1}), which requires $\phi^{-1}c$ to be small. While it is small
at the glass transition, $\phi^{-1}c = s c \mathcal{N}$ may not be
small at the higher temperatures considered.  The right-hand side of
BB\&T's formula is missing the requisite factor of $(1-\phi^{-1}c)$
present in Eq.\ (\ref{C1}).  It is also missing a possible temperature
dependence of $\phi$.  This temperature dependence would reflect that
of transport in a liquid.  It is typically more pronounced for
isobaric temperature variation than for isochoric temperature
variation, but weak and sub-Arrhenius in either
case~\cite{Jonas}. Thus, we write $\phi \approx
\phi_{\mathrm{g}}+\phi_{1}\left( T-T_{\mathrm{g}}\right)/T_g $, where
the parameters $\phi_{\mathrm{g}}$ and $\phi_{1}$ are temperature
independent.

In Ref.~\cite{JPGDC03}, we chose to not consider $C$ above $T_g$
because doing so requires the additional parameters described in the
previous paragraph. But since, BB\&T raise this point, we show in
Fig.1 that the effects of these extra factors on the temperature
dependence of $C$ is significant. This figure can be contrasted with
that in BB\&T's~\cite{BBT}.  While the asymptotic formula predicts an
exponential variation with temperature, the full formula predicts
modest variation with temperature that is not inconsistent with
experiment.  We also show fits to the corresponding excess entropy,
$S_{\mathrm c}$.  Better agreement between theory and experiment for
these properties would require modification of the models and perhaps
further parameterization.

\bigskip

\acknowledgments

In the US, this work was supported initially by the NSF, and more
recently by DOE grant no.\ DE-FE-FG03-87ER13793. In the UK, it was
supported by EPSRC grants no.\ GR/R83712/01 and GR/S54074/01, and
University of Nottingham grant no.\ FEF 3024.

\end{document}